\documentclass{TTP_DSL2006}

\usepackage[dvipsone]{graphicx}
\usepackage[intlimits]{amsmath}
\usepackage{amssymb}
\usepackage{exscale}

\journalcitation{Materials Science Forum}

\copyrightyear{2006}

\begin{document}

\title{Electronic structure of graphite/6H-SiC interfaces}

\author{Th. Seyller\inst{1}$^{,\,\text{a}}$, K. V. Emtsev\inst{1}, F. Speck\inst{1}, K.-Y. Gao\inst{1}, L. Ley\inst{1}}

\institute{Lehrstuhl f\"ur Technische Physik, Universit\"at Erlangen-N\"urnberg\\Erwin-Rommel-Str. 1, D-91058 Erlangen, Germany}
\maketitle

\sffamily
\begin{center}
 $^\text{a}$thomas.seyller@physik.uni-erlangen.de
\end{center}

\vspace{3mm} 
\hspace{-7.7mm} 

\baselinestretch
\renewcommand{\baselinestretch}{0.95}

\vspace{3mm} 
\hspace{-7.7mm}
\rmfamily
\normalsize
\textbf{Abstract.} We have studied the electronic structure of the interface between 6H-SiC\{0001\} and graphite. On n-type and p-type 6H-SiC(0001) we observe Schottky barriers of $\phi_{b,n}^{Si}$ = 0.3$\pm$0.1 eV and  $\phi_{b,p}^{Si}$ = 2.7$\pm$0.1~eV, respectively. The observed barrier is face specific: on n-type 6H-SiC($000\overline{1}$) we find $\phi_{b,n}^{C}$ = 1.3$\pm$0.1 eV. The impact of these barriers on the electrical properties of metal/SiC contacts is discussed.

\section{\label{sec:introduction}Introduction}

The formation of contacts is an important technological aspect of device fabrication. Ohmic contacts on SiC require a post deposition anneal (PDA) of the deposited metal film. PDA induces a chemical reaction between the metal and SiC. Metal silicides are frequently observed (see refs. \cite{crofton1997a,tanimoto2003a} and references therein) after PDA. For Ni \cite{crofton1995a} and Co \cite{lundberg1993a} graphite was observed as a byproduct of the silicide formation. In order to elucidate the impact of graphite on the electrical properties of metal contacts we have determined the Schottky barriers between graphite and 6H-SiC\{0001\} surfaces. Both, (0001) and ($000\overline{1}$) oriented surfaces have to be considered for devices such as Schottky diodes, pn-diodes, MOSFETs and JFETs. 

\section{\label{sec:experimental}Experimental}

The samples used in this study were on-axis oriented, n-type 6H-SiC(0001) with a doping level of around $1\times10^{18}$~cm$^{-3}$, 3.5 degree off axis oriented, p-type 6H-SiC(0001) with a p-type epilayer with a doping concentration of $1\times10^{16}$~cm$^{-3}$, and  on-axis oriented, n-type 6H-SiC$(000\overline{1})$ with a doping level of approximately $1\times10^{18}$~cm$^{-3}$.

Graphite layers were grown by solid state graphitization \cite{forbeaux1998a,forbeaux1999a}, i.e. by sublimation of Si from the surface at elevated temperatures ($\geq1150^{\circ}$C) as described elsewhere \cite{seyller2006a,seyller2006b,emtsev2006b}. First the samples were annealed in a Si flux at 950$^{\circ}$C. Thereafter, Si was gradually removed from the surface by annealing steps at increasing temperatures until graphitization occured. The sequence of surface reconstructions observed during the preparation is face-dependent \cite{starke2003a}. On SiC(0001) it is $(3\times3)$, $(\sqrt3\times\sqrt3)R30^\circ$, $(6\sqrt3\times6\sqrt3)R30^\circ$, and $(1\times1)_{\mathrm{graph}}$. On SiC($000\overline{1}$) different reconstructions occur: $(2\times2)_{\mathrm{Si}}$, $(3\times3)$, $(2\times2)_{\mathrm{C}}$, $(1\times1)_{\mathrm{graph}}$. In agreement with previous results \cite{forbeaux1998a,forbeaux1999a} the graphite layers obtained on 6H-SiC(0001) showed a sharp LEED pattern which indicated that the graphite lattice was rotated against the substrate by 30 degrees \cite{seyller2006a,seyller2006b}. In contrast to that, graphite layers grown on 6H-SiC($000\overline{1}$) led to a ring-like LEED pattern indicative of rotationally disordered domains. The rings had some intensity modulation with maxima in positions 30 degrees from the substrate spots suggesting a preference for an alignment of the graphite overlayer.

Photoelectron spectroscopy measurements were carried out using the end station MUSTANG operated either at beam line U49/2-PGM1 or at U49/2-PGM2 of the synchrotron radiation source BESSY II. The end station is equipped with a Specs Phoibos 150 electron analyzer. The combined energy resolution was estimated from the width of the Fermi level measured of the Mo-made sample holder and ranged from 60~meV at $\hbar\omega=130$~eV to 150~meV at $\hbar\omega=700$~eV.
 
\section{\label{sec:results}Results}

\begin{figure}
    \centering
    \includegraphics [width=6cm]{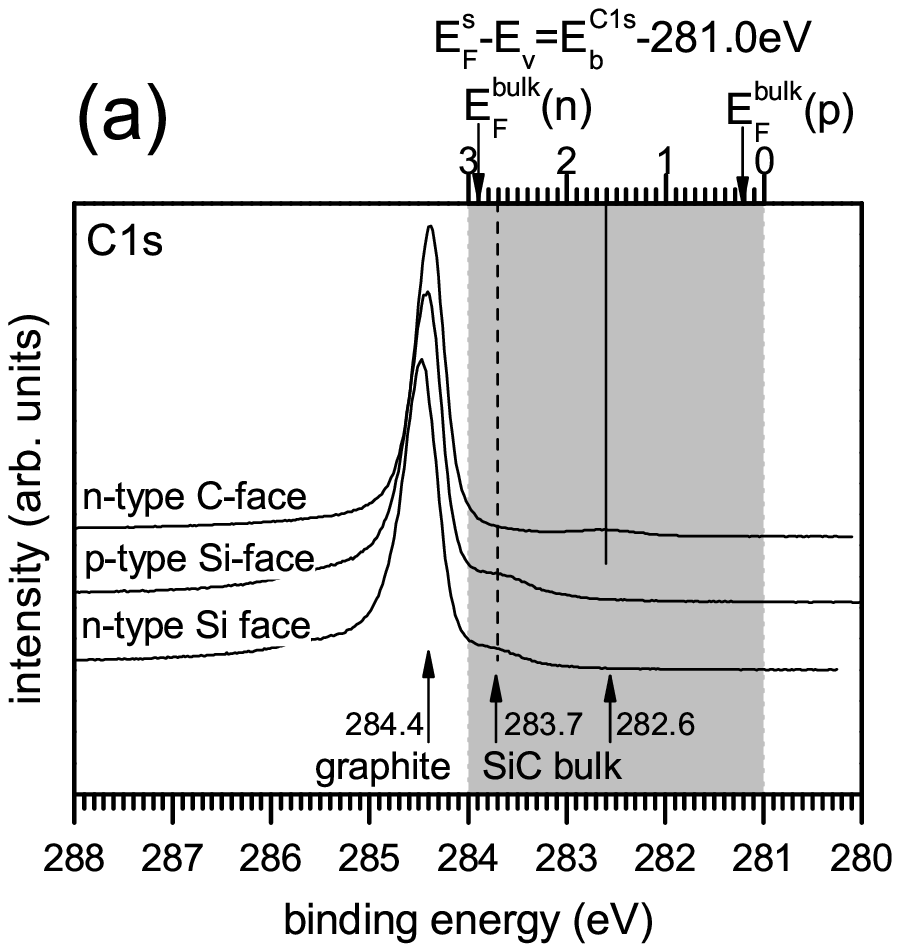}
    \includegraphics [width=6cm]{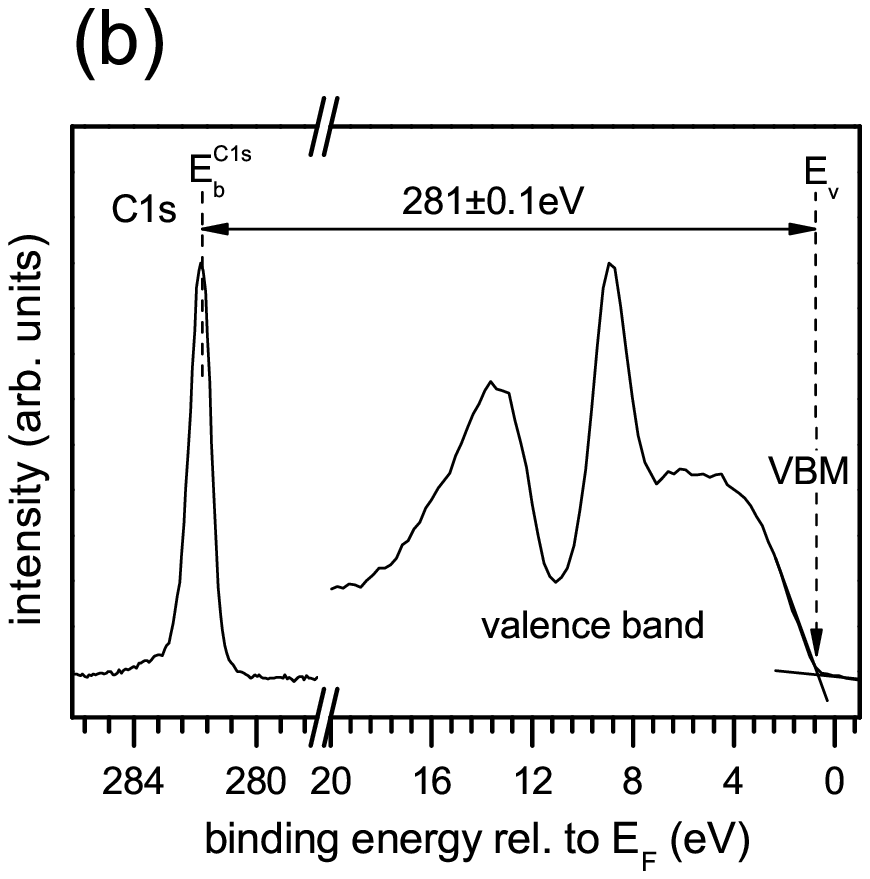}
    \includegraphics [width=4cm]{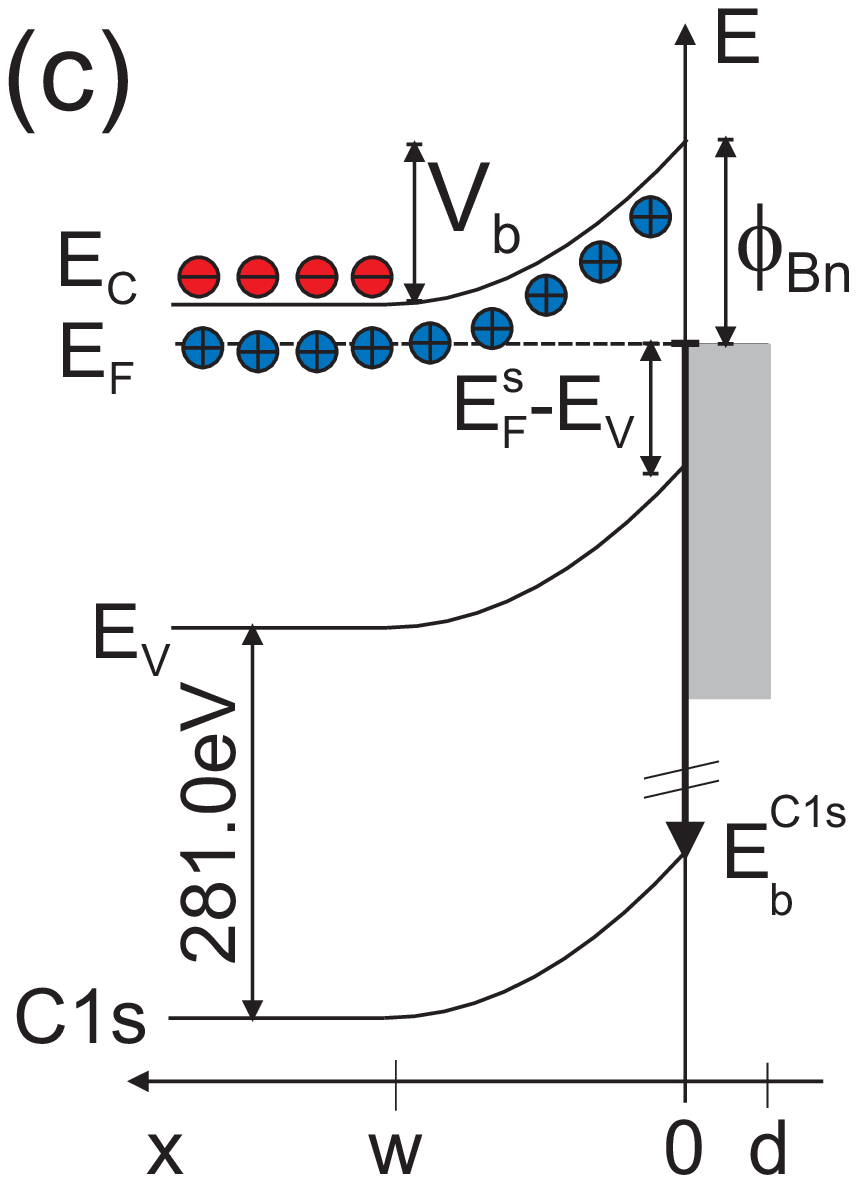}
    \caption{(a) C~1s core level spectra of 2~nm graphite grown on n-type 6H-SiC(0001), p-type 6H-SiC(0001), and n-type 6H-SiC($000\overline{1}$) by solid state graphitization. (b) Determination of the energy difference between the C~1s core level and the valence band maximum with XPS.(c) Band diagram of the n-type SiC/graphite contact.}
    \label{fig:C1sspectra}
\end{figure}

Fig. \ref{fig:C1sspectra}(a) displays C~1s core level spectra measured for 2~nm thick graphite films on the three different samples mentioned above. The overlayer leads to a sharp, asymmetric line at 284.41$\pm$0.06 eV, which is typical for graphite. In addition, weak signals of the underlying SiC substrate are visible as well. For the Si-face samples the binding energy of the SiC substrate was practically the same $E_b^{C1s}=283.7\pm0.06$~eV \cite{seyller2006b} independent of the doping type (see  table \ref{tab:properties}). On the n-type C-face sample the binding energy was $E_b^{C1s}=282.6\pm0.06$~eV.

In order to determine the Schottky barrier from the measured core levels we first determine the difference between binding energy $E_b^{C1s}$ of the bulk C1s core level and the valence band maximum $E_v$ for SiC as shown in figure \ref{fig:C1sspectra}(b). This quantity amounts to 281.1$\pm$0.1~eV. Note, that the binding energies determined by photoelectron spectroscopy are referenced to the Fermi level (see figure \ref{fig:C1sspectra}(c)). The inelastic mean free path of photoelectrons is short (typically $\leq$~2~nm for the present experiments) compared to the width of the depletion layer $w$ (see below), i.e. we are probing only a small region close to the surface in which the band bending can be considered constant. The position of the Fermi level at the surface $E_F^s$ with respect to the valence band maximum is thus given by $E_F^s-E_v=E_b^{C1s}-281.0$~eV. The extracted numerical values are collected in table \ref{tab:properties}. The position of the bulk Fermi level with respect to the valence band maximum $E_F^{bulk}-E_v$ estimated \cite{sze1981} for our n-type and p-type 6H-SiC is also listed. The band bending $V_b$ present at the surface of SiC is simply the difference between the bulk and the surface Fermi level, i.e. $e_0V_b=E_F^s-E_F^{bulk}$. From the band bending we have also calculated the surface charge $Q_{sc}$ and the width of the depletion layer $w$ (see table \ref{tab:properties}). In all cases studied here, the latter is much larger than the inelastic mean free path of the electrons, indeed. Finally table \ref{tab:properties} also lists the derived Schottky barriers which are $\phi_{b,p}=E_F^s-E_v$ and $\phi_{b,n}=E_g-(E_F^s-E_v)$ for the case of p-type and n-type SiC, respectively. For the Si-face the Schottky barriers are dramatically different: quite small ($\phi_{b,n}^{Si}$=0.3$\pm$0.1~eV) for electrons and huge ($\phi_{b,p}^{Si}$=2.7$\pm$0.1~eV) for holes \cite{seyller2006b}. On the other hand, on the C-face a sizable barrier of $\phi_{b,n}^{C}$=1.4$\pm$0.1~eV for electrons was derived. In addition, taking into account that $\phi_{b,n}+\phi_{b,p}=E_g$ we predict that $\phi_{b,p}^{C}$=1.6$\pm$0.1~eV. 
 
\begin{table}
\begin{center}
\vspace{4mm}
\small
\begin{tabular}{c|c|c|c|c|c|c|c|c} 
  Surface & Doping & $E_b^{C1s}$ & $E_F^s-E_v$ & $E_F^{bulk}-E_v$ & $V_b$ & $Q_{sc}$ & $w$ & $\phi$ \\
   & (cm$^{-3}$) & (eV) & (eV) & (eV) & (eV) & ($e_0$) & (nm) & (eV) \\  
  \hline\hline
  (0001) & N, $1\times10^{18}$ & 283.7 & 2.7 & 2.9 & 0.2 & $-9.5\times10^{11}$ & 23 & 0.3$\pm$0.1\\
  (0001) & Al, $1\times10^{16}$ & 283.7 & 2.7 & 0.2 & -2.5 & $2.9\times10^{11}$ & 958 & 2.7$\pm$0.1 \\
  \hline
  $(000\overline{1})$ & N, $1\times10^{18}$ & 282.6 & 1.6 & 2.9 & 1.3 & $-2.5\times10^{12}$ & 58 & 1.4$\pm$0.1 \\
\end{tabular}\\
\normalsize
\caption{\label{tab:properties}C~1s binding energy of SiC ($E_b^{C1s}$), position of the Fermi level above the valence band maximum at the surface ($E_F^s-E_v$) and in the bulk ($E_F^{bulk}-E_v$), band bending ($V_b$), surface charge ($Q_{sc}$), width of depletion region ($w$), and barrier ($\phi$) of the graphite/SiC interfaces for graphite grown on n-type and p-type 6H-SiC(0001), and n-type 6H-SiC($000\overline{1}$), respectively.}
\end{center}
\end{table}

\section{\label{sec:discussion}Discussion}

On the Si-face we observe a dramatic difference between the barriers for holes and electrons. For n-type 6H-SiC(0001) the barrier is extremely low. Thus, Schottky contacts on n-type SiC(0001) are expected to degrade if graphite is formed at the interface. Indeed this is observed experimentally: Co/SiC Schottky contacts transform into ohmic contacts after annealing at 900~$^{\circ}$C \cite{lundberg1993a}. The transformation is accompanied by the formation of Co$_2$Si and a buildup of carbon in the contact layer. Consequently, for Schottky contacts to n-type SiC(0001) we recommend metals for which the phase diagram prohibits the formation of graphite.

On the other hand, ohmic contacts on this surface orientation could benefit from interfacial graphite. In fact, Lu et al. \cite{lu2003a} observed an ohmic behavior of pure carbon contacts on n-type 4H-SiC(0001) annealed at 1150-1300~$^{\circ}$C which is consistent with a small Schottky barrier between n-type SiC(0001) and graphite. In another work, Lu et al. \cite{lu2003b} observed that Ni/carbon/SiC(0001) and Co/carbon/SiC(0001) stacks annealed at 800~$^{\circ}$C developed ohmic behavior but similar stacks with W, Mo, Au, or Al remained rectifying. They also reported that the ohmic contacts contained graphite while the rectifying did not. According to our results, the observed contact behavior is readily explained by the small value for $\phi_{b,n}^{Si}$ for the graphite/SiC(0001).

Our results suggest that graphite will be detrimental for ohmic contacts to p-type SiC(0001) and that on p-type SiC(0001) metals should be preferred which form stable carbides or which have a high solubility for carbon. In agreement with that, Al/Ti contacts have shown low contact resistivities (see ref. \cite{tanimoto2003a} and references therein). Furthermore, metal/Si co-deposition is expected to lead to better ohmic contacts than simple deposition of metals. This is so because in a metal/Si/SiC stack, silicide formation can occur without SiC decomposition and graphite formation. Accordingly, Lundberg and \"Ostling \cite{lundberg1996a} reported that Si/Co/SiC stacks annealed at 900~$^{\circ}$C showed an ohmic behavior superior to Co/SiC contacts treated in the same way. 

Our experiments also show a clear-cut face dependence of the barriers. The microscopic origin of the different barriers observed on Si-face and C-face is unclear as yet. Considering that different reconstructions occur on SiC(0001) and ($000\overline{1}$), it is very likely that the atomic arrangement of the interface between graphite and these two polar surfaces is not the same. On SiC(0001) the interface consists of the $(6\sqrt3\times6\sqrt3)R30^\circ$ structure \cite{emtsev2006b}. The atomic arrangement at the interface between SiC($000\overline{1}$) and graphite is currently unknown. A different interface structure on the atomic scale is likely to produce different interface dipole moments as well as different interface states which may pin the Fermi level. In terms of contact properties we note that the large value of $\phi_{b,n}^{C}$=1.4$\pm$0.1~eV observed here supports the conclusions of Nikitina et al. \cite{nikitina2005a}, namely, that graphite is not responsible for ohmic contact behavior of Ni/SiC($000\overline{1}$) after PDA.

\section{\label{sec:conclusion}Conclusion}

We have determined the Schottky barriers between graphite and n- and p-type 6H-SiC(0001) as well as n-type 6H-SiC($000\overline{1}$). While we find a rather small value of $\phi_{b,n}^{Si}$=0.3$\pm$0.1~eV on n-type 6H-SiC(0001), the Schottky barrier on p-type SiC(0001) is extremely large $\phi_{b,p}^{Si}$=2.7$\pm$0.1~eV. The barrier is strongly face dependent: for n-type 6H-SiC($000\overline{1}$) we determined $\phi_{b,n}^{C}$=1.4$\pm$0.1~eV.  The origin for the face dependence is the subject of ongoing investigations. Our results give a consistent interpretation of numerous reports on the contact behavior of metals on SiC\{0001\} after PDA.

\section{Acknowledgments}

The authors are grateful for support by the BESSY staff and by G. Gavrila from TU Chemnitz. Traveling costs for beam times were provided by the BMBF through Grant No. 05 ES3XBA/5.


\end{document}